# A VLA SEARCH FOR RADIO SIGNALS FROM M31 AND M33

ROBERT H. GRAY
Gray Consulting, 3071 Palmer Square, Chicago, IL 60647, USA
Email: RobertHansenGray@gmail.com

KUNAL MOOLEY
Oxford Centre for Astrophysical Surveys, Denys Wilkinson Building, Keble Road, Oxford, OX1 3RH  England
Email: kunal.mooley@physics.ox.ac.uk

## ABSTRACT

Observing nearby galaxies would facilitate the search for artificial radio signals by sampling many billions of stars simultaneously, but few efforts have been made to exploit this opportunity.  An added attraction is that the Milky Way is the second-largest member of the Local Group, so our galaxy might be a probable target for hypothetical broadcasters in nearby galaxies.  We present the first relatively high spectral resolution (<1 kHz) 21-cm band search for intelligent radio signals of complete galaxies in the Local Group with the Jansky VLA, observing the galaxies M31 (Andromeda) and M33 (Triangulum)—the first and third largest members of the group respectively—sampling more stars than any prior search of this kind.  We used 122 Hz channels over a 1 MHz spectral window in the target galaxy velocity frame of reference, and 15 Hz channels over a 125 kHz window in our local standard of rest.  No narrowband signals were detected above a signal-to-noise ratio of 7, suggesting the absence of continuous narrowband flux greater than approximately 0.24 Jy and 1.33 Jy in the respective spectral windows illuminating our part of the Milky Way during our observations in December 2014 and January 2015.  This is also the first study in which the upgraded VLA has been used for SETI.

Key words: astrobiology, extraterrestrial intelligence, galaxies: individual (M31, M33)

## 1. INTRODUCTION

### 1.1 The Search for Extraterrestrial Intelligence

The search for extraterrestrial intelligence (Tarter 2001) or SETI searches for evidence of life elsewhere than on the Earth, often by looking for evidence of technological activity such as radio (Cocconi & Morrison 1959) or optical (Schwartz & Townes 1961) signals.  It's unknown if life exists elsewhere, or how often it might be intelligent and produce detectable signals, but the possibility of electromagnetic signaling permits searching on a very large scale.  Many other search strategies are possible (Cabrol 2016) such as looking biosignatures in the atmospheres of exoplanets transiting their stars (Ehrenreich et al. 2006), but signaling has a much larger potential range.  Radio signals can span our entire galaxy and reach between galaxies, allowing an enormous number of stars to be sampled, and observing many stars presumably improves the chances of finding one of interest.  This article reports relatively brief (5-20 min.) radio searches of relatively narrow spectral windows (0.125-1 MHz) covering ~$10^{12}$ stars in two of our largest neighboring galaxies, which may include more stars than ever sampled by SETI experiments before.

### 1.2  The Problem of Direction

A fundamental problem in SETI is the large number of directions or objects that may need to be searched if high-gain antennas are used to achieve high sensitivity (Ekers 2002).  For example, a 30 m diameter antenna system with a 0.5º beamwidth at a wavelength of 21 cm must be pointed in ~$10^5$ different directions to tessellate the sky, and a 300-m has ~$10^7$ directions.  Total search time increases with the time spent pointing in each direction—integrating or searching in frequency or searching in time for intermittent signals or other activity.  Larger and more sensitive telescopes can increase search speed if sensitivity is the only criteria (requiring less integration time for a given sensitivity), but if we also wish to search in time for possibly intermittent signals we need to dwell for some constant time in each direction.  Dwelling for five minutes in each direction—implicitly assuming signals are present all or much of the time—would take a year to cover the sky using a 30-m antenna system, and a century using a 300-m.  Dwelling for 24 hours in each direction—possibly searching for low-duty-cycle signals—would take decades with a 30-m and millennia with a 300-m.  Extremely long search times are not practical.

## 1.3. Rationale for Searching Nearby Galaxies

One strategy for increasing search speed is to observe large concentrations of stars and therefore presumably planets (Petigura et al. 2013) in order to include more potential signal sources in each pointing. The distribution of stars in our galaxy is approximately uniform out ~$10^3$ ly (Ekers 2002) but further out, stars are concentrated in the plane (a light year is about 0.3 parsec and is commonly used in SETI discussions; one advantage is that it gives signal propagation time). Searches in limited parts of the Milky Way plane have been carried out for narrowband ~1 Hz radio signals (Backus 2005) as well as wideband ~100 kHz emissions from astrophysical transients (Williams 2013), for several examples.

Some galaxies in the Local Group offer even larger concentrations of stars. M31 has ~$10^{12}$ stars in approximately 3 deg$^2$ and M33 has ~$10^{10}$ stars in 1 deg$^2$; for comparison, our galaxy is thought to contain ~$10^{11}$ stars, with an angular size of a few square degrees seen from those two galaxies (number of stars is typically estimated from mass and not based on actual counts).

A 30-m antenna system can observe either M31 or M33 with only a handful of pointings at 21 cm, compared with ~$10^5$ pointings to survey the entire sky. Roughly speaking, we could spend a year searching ~$10^{11}$ stars in the Milky Way with a hundred thousand pointings, or we could search ~$10^{12}$ stars in M31 with a handful of pointings in a matter of hours. This seems like a strong reason to search M31, and a similar case can be made for M33.

Another reason to search other galaxies in the Local Group is that broadcasters located in the group would have some reason to transmit toward the Milky Way because it is the second-largest galaxy in the group. And, astronomers in the Milky Way have a reason to point high-gain antenna systems toward nearby galaxies such as M31 and M33—to study their structure and kinematics, especially at the 21 cm wavelength of neutral hydrogen—even if they are not looking for artificial radio signals.

Another reason to search nearby galaxies is that we can dispense with some assumptions often implicit in SETI, such as an isotropic broadcast (requiring a great deal of power for even modest range), or a targeted broadcast directed at our solar system (requiring that we be selected as a target). In the case of signals from other galaxies, the entire Milky Way might be illuminated (requiring a great deal of power, but not an isotropic broadcast), or more highly directive broadcasts might illuminate sectors of the Milky Way (but seem unlikely to be directed at our specific solar system due to the range). An assumption implicit in targeted searches of nearby stars is that the number of planets with life, intelligence, and broadcasters in our galaxy must be large in order to have a good chance of any existing nearby, while searching other galaxies allows a minimum assumption about the number—as few as one broadcaster in a galaxy.

Another reason to search other galaxies is the so-called Fermi paradox, an argument that no other cases of technological intelligence exist in the Milky Way galaxy. The argument "they are not here; therefore they do not exist" in our galaxy (Hart 1975) appears to be the origin of the so-called Fermi paradox, although it does not appear not to have been Fermi's view and is not a logical paradox (Gray 2015). Hart assumed that interstellar travel and colonization would fill the Milky Way in a small fraction of its age, so we should see evidence of other technological intelligence on Earth if it existed anywhere else in our galaxy—and he concluded that it must not exist, so searching elsewhere within our galaxy is pointless. Even if this is viewed as a strong argument, it would have little power with respect to other galaxies.

We used the Jansky VLA to search for radio signals in M31 and M33 because its relatively high sensitivity is appropriate for the long range, its WIDAR spectrometer offers relatively high spectral resolution, and its synthesis imaging capability helps discriminate against interference. This is the first reported search of M31 and M33 for artificial radio signals, although several unpublished efforts are noted later. Covering ~$10^{12}$ stars, it sampled more stars than any previous published search. SETI observations using the full VLA were carried out on three occasions in the past (Gray & Marvel 2001; Shirai et al. 2004); our search is the first of its kind with the upgraded VLA.

## 1.4. Drawbacks

One problem with signaling between galaxies would be the large range, implying a large amount of power for transmission. M31 and M33 are approximately $2.5 \times 10^6$ and $2.6 \times 10^6$ ly distant respectively (McConnachie et al. 2005), compared with a $5 \times 10^4$ ly distance scale in the Milky Way (its radius)—a factor of about 50 larger, and a factor of 2500 in power since power required increases with the square of range.

The power required for transmission is:

$$P_t = (4 \pi S R^2) / G_t$$

where $P_t$ is power in watts, $S$ is sensitivity of the receiver in W/m$^2$, $R$ is the range in meters, and $G_t$ is the gain of the transmitting antenna system $(4 \pi A_{eff}) / \lambda^2$ where $A_{eff}$ is the effective area and $\lambda$ is wavelength both in meters. Sensitivity of a receiver is (following Gulkis, Olsen & Tarter 1979):

$$S = [(4 \; SNR \; k \; T_{sys}) / (\pi \; R_{eff} \; D_r^2)] \; (b/t)^{0.5}$$

where SNR is desired signal-to-noise ratio, k is Boltzmann's constant $1.38 \times 10^{-23}$ J/K, $T_{sys}$ is the receiver system temperature in Kelvins, $R_{eff}$ is receiver antenna efficiency, $D_r$ is receiver antenna diameter in meters, b is receiver bandwidth in Hz, and t is integration time in seconds.

Table 1 shows the power required for signaling under several hypothetical scenarios discussed below, assuming a wavelength of 21 cm.

| Table 1 Power Required (W) for Detection at $2.5 \times 10^6$ ly Range | | | |
|---|---|---|---|
| | Transmitter antenna diameter (m) | | |
| | 10 | 100 GBT-scale | 1,000 SKA-scale |
| Pointings to cover Milky Way (~3 deg$^2$) | ~1 | ~$10^4$ | ~$10^6$ |
| Receiver System | | | |
| JVLA: b=122 Hz, t=20 min. | $6 \times 10^{16}$ | $6 \times 10^{14}$ | $6 \times 10^{12}$ |
| JVLA: b=15 Hz, t=5 min. | $4 \times 10^{16}$ | $4 \times 10^{14}$ | $4 \times 10^{12}$ |
| SKA-scale: b=1 Hz, t=1 sec., $R_{eff}=T_{eff}=1.0$, $T_{sys}=20$ | | | $8 \times 10^{10}$ |
| Note: λ=21 cm, SNR=7 in all examples. | | | |

Our search could detect considerably less than $10^{17}$ W radiated from M31 or M33 if it was radiated by an antenna system with an approximately 3° beamwidth illuminating the entire Milky Way galaxy (single-element antennas are used in examples for gain and beamwidth calculations, but would not be practical for so much power). That is much more power than current terrestrial energy consumption of ~$10^{13}$ W and is on the scale of terrestrial solar insolation of ~$10^{16}$ W.

But, more advanced civilizations (or comparable entities) might have access to much more power. In one hypothetical classification scheme (Kardeshev 1964, 1967), a Type I civilization would use $10^{12}$ W, the terrestrial production around 1964, although Type I is sometimes generalized to our $10^{16}$ W solar insolation (Lemarchand 1992). A Type II would be capable of harnessing the $10^{26}$ W output of a star like our Sun, perhaps using some variation on a Dyson sphere (Dyson 1960), and a Type III would possess energy on the $10^{37}$ W scale of a galaxy like ours. The power needed to produce a detectable signal in our search would be a tiny fraction of the power available to a Type II or III civilization. In the context of SETI, $10^{17}$ W is comparable to the power required for an isotropic broadcast with a range of 10,000 ly inside the Milky Way, assuming a 300-m receiver antenna, 1 Hz channels, and SNR=7.

A second example illustrates the dramatic reduction in power required if a higher-gain antenna system is used for transmission, illuminating parts of a galaxy in succession, although duty cycle and therefore search time then become factors in detection. A 100-m antenna system (Green Bank Telescope scale) broadcasting from M31 or M33 might tessellate the Milky Way into some $10^4$ beam areas, and if it illuminated each area for 5 min. in succession with ~$10^{15}$ W, our search could detect the signal if it was present when we looked—although it might illuminate our area only once per month. Larger broadcast antennas would reduce power requirements to familiar levels; our search could detect a ~$10^{13}$ W broadcast from a 1,000-m antenna system (Square Kilometer Array scale), and ~$10^{11}$ W from a 10,000-m antenna system, if they were pointed our way during our observations.

A final example illustrates that communication between galaxies might not greatly exceed the scale of present-day terrestrial scientific projects, if both parties know the other's location. In that case, both could use very high-gain antenna systems, greatly reducing power requirements. With a 1,000-m antenna system on each end, approximately $10^{11}$ W would suffice for transmission, which is five times the power generated by the 22.5 GW Three Gorges Dam in China and seven times the 14 GW Itaipu Dam between Brazil and Paraguay. Building those dams cost several times $10 bn, and taking the cost of a SKA as $1 bn, the capital cost of one end of the link would be roughly comparable to the ~$100 bn cost of the International Space Station.

Other drawbacks to signaling between galaxies include the very long propagation time and questions such as what might motivate and sustain such long-term activities, but none of these considerations seem so severe as to rule out the possibility that observations might find something interesting.

1.5. Prior Galaxy Observations

Radio astronomers have observed M31 and M33 many times and have not reported any artificial radio signals (for example Corbelli et al. 2010; Gratier et al. 2010; Thilker et al. 2002; Dickey & Brinks 1993; Braun 1990; Deul & van der Hulst 1987; Brinks & Shane 1984). But, such observations are not intended to detect narrowband radio signals. They typically use ~10 kHz channels (2 km s$^{-1}$ at 21 cm), while radio signals might be much narrower. Signals 1 Hz wide—a channel width often mentioned in the context of interstellar communication (e.g. Oliver & Billingham 1971, p. 31)—would be attenuated by a factor of $10^4$ in a 10 kHz channel. It is also common practice in radio astronomy to flag and ignore obvious or intermittent narrow bandwidth signals as RFI, which is very often the case.

A review of SETI observations (Tarter 1995 and updates) found only a few that observed galaxies with high spectral resolution (≤1 kHz) and high sensitivity, all unpublished.

Sagan and Drake observed four galaxies using Arecibo with 1 kHz resolution for four hours (mentioned in Sagan & Drake 1975), including 212 positions at M33 (Morrison et al. 1977) dwelling for about 60 sec. on each, but M31 was not observed because it is outside Arecibo's declination range (RG personal communication with Drake, F. D., August 28, 2015).

M31 and M33 were observed in 1990 by Gray for five hours/beam using the Harvard/META radio telescope (Horowitz & Sagan 1993) with 0.05 Hz resolution in several 400 kHz windows in the 21 cm band (a width of 80 km s$^{-1}$) Doppler-corrected to several velocity frames including the LSR and CMB, but not corrected for the relatively large target galaxy velocity—up to 600 km s$^{-1}$ or 3 MHz for M31 (Brinks & Shane 1984) and up to 300 km s$^{-1}$ for M33 (Deul & van der Hulst 1987; Putman et al. 2009) so probably would have missed 21-cm band signals in the galaxy's velocity frame of reference. No results were published because very little data could be recorded for analysis.

A survey for artificial signals from the Small Magellanic Cloud has been reported (Shostak et al. 1996), observing three 14 arcmin fields with 1 Hz resolution from 1.2-1.75 GHz integrating for 130 s. The number of stars observed was reported as >$10^7$ stars and the area observed was approximately 1% of the total area.

The Ohio State transit survey (Dixon 1985) and the META transit survey (Horowitz & Sagan 1993) would have swept across both M31 and M33 and many other galaxies for several minutes several times during the course of their surveys with 10 kHz and 0.05 Hz resolution respectively, but the respective 500 kHz and 400 kHz spectral windows were not Doppler-corrected for the velocity of those galaxies because they were not explicit targets.

No galaxy was listed as a target of the Allen Telescope Array SETI program (setiQuest Data Links 2013) although the ATA has observed M31 and M33 in the hydrogen band using wide (>10 kHz) channels (Welch et al. 2009). Some searches of other galaxies for infrared evidence of large-scale astro-engineering have been made (Wright et al. 2014, for example).

Many searches for narrowband radio signals have been reported, but with caveats noted above, none have been reported for these Local Group galaxies at high spectral resolution.

## 2. OBSERVATIONS

2.1 Fields Observed

Five fields were observed along the major axis of M31 and three fields along the major axis of M33 as illustrated in Figure 1 showing 0.5° FWHP circles superimposed on POSS II (Reid et al. 1991) optical images. Some fields were overlapped to get longer total observing time and to allow the possibility of multiple detections. Most data analysis imaged approximately 1° fields, covering more area but with reduced sensitivity past the half-power points.

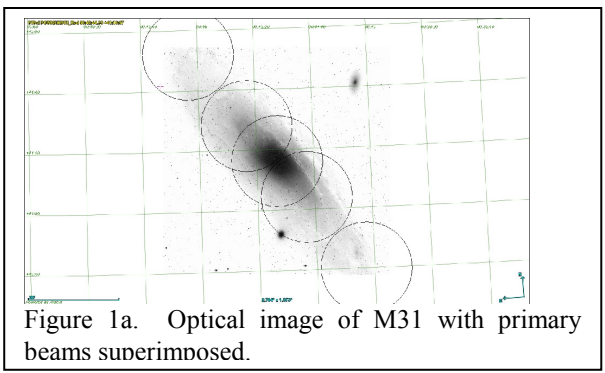
Figure 1a. Optical image of M31 with primary beams superimposed.

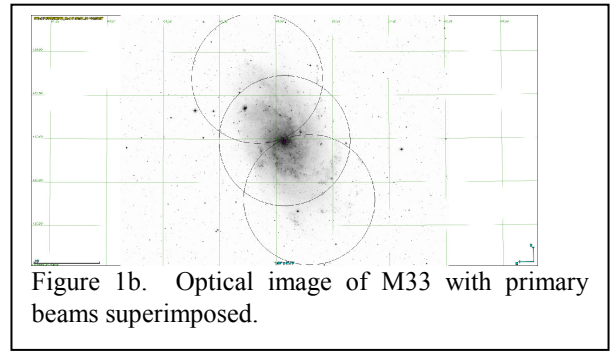
Figure 1b. Optical image of M33 with primary beams superimposed.

Coordinates of field centers are in Table 2.

| Table 2 Target Coordinates | | |
|---|---|---|
| Field name | R.A. (J2000.0) | Decl. (J2000.0) |
| M31 ... N3 | 00 45 16 | 41 52 00 |
| M31 ... N1 | 00 43 36 | 41 28 00 |
| M31 ... CTR | 00 42 44 | 41 16 09 |
| M31 ... S1 | 00 41 54 | 41 04 00 |
| M31 ... S3 | 00 40 14 | 40 40 00 |
| M33 ... N1 | 01 34 20 | 30 54 00 |
| M33 ... CTR | 01 33 51 | 30 39 37 |
| M33 ... S1 | 01 33 25 | 30 26 00 |
| VGR1 | 17 11 58 | 11 58 05 |
| Note: Units of right ascension are hours, minutes, and seconds, and units of declination are degrees, arcminutes, and arcseconds. | | |

2.2. Telescope Description

The telescope is described in Table 3. Some observations were made during the change in array configuration from C to CnB and from CnB to B, which resulted in spatial resolution varying from 14" to 4.3" for 21-cm observations.

| Table 3 Telescope Description | |
|---|---|
| Telescope | Jansky Very Large Array |
| Antenna system | 27 x 25-m element interferometer |
| Primary FWHP | 31.7' (1.4 GHz); 5.4' (8.4 GHz) |
| Spatial resolution | 14" (C config.); 4.3" (CnB) at 21 cm |
| Channels | 8192 |
| Channel width | 15.3 Hz, 122 Hz, 1.95 kHz |
| Polarizations | 2 circular |
| Integration time | 5 sec. |
| System temperature | 35 K (L-band); 34 K (X-band) |

2.3. Spectral Window Selection

Ideally, searches for interstellar signals would monitor much of the electromagnetic spectrum in all directions simultaneously with high spatial and spectral resolution and high sensitivity, but that is not currently practical, so choices of direction, spectral window, and dwell time must be made.

In the case of signaling between nearby galaxies, both hypothetical broadcasters and searchers know each other's location within a few square degrees, dramatically reducing the number of directions that need to be searched in contrast with all-sky surveys. Both presumably also know that astronomers sometimes point high-gain antennas with spectrometers toward neighboring galaxies to study their structure and kinematics in the hydrogen band, which is one reason broadcasters might transmit near the wavelength of neutral hydrogen, and for searchers to observe that band at high resolution.

We selected the 21 cm band as particularly appropriate for a galaxy search because both broadcasters and searchers would be aware of its use in radio astronomy, and for several additional reasons. One reason is that the 21-cm band has been suggested for interstellar communication (Cocconi & Morrison 1959) and that band has been selected in major searches (e.g. Dixon 1985; Horowitz & Sagan 1993). Another reason is that the band might be generally protected for radio astronomy observations, as it is on Earth. A hypothetical broadcaster considering the unknown aggregate radio spectrum across many hypothetical searchers might anticipate that much of the spectrum would be occupied by local emissions, but have notches at wavelengths useful for radio astronomy such as HI.

We were able to search only a limited subset of parameter space at high spectral resolution—1 MHz and 0.0125 MHz at 21 cm, which are relatively narrow spectral windows—but arguably the best spectral windows under the constraint of limited bandwidth.

As summarized earlier, the JVLA (Perley et al. 2011) is well suited for this search for a number of reasons. First, its high sensitivity is appropriate for the long range, and it can view both M31 and M33, while the larger Arecibo telescope can not view M31. Second, an imaging interferometer like the JVLA helps discriminate against radio frequency interference, which does not in general map to a point source on the sky (Thompson 1982, Bridle 1994). Finally, the WIDAR correlator supports many channels (e.g. 8192 in two circular polarizations) and channels as narrow as 15 Hz are practical before calibration times become excessively long (much more than 50% overhead). Synthesis imaging allows many potential sources of signals to be observed simultaneously, as opposed to observing single targets one-at-a-time with a high-gain antenna which is a slow process.

The VLA's high spatial resolution can sometimes identify the optical counterpart of a radio source, and a single-channel point-like source very near the coordinates of a star could suggest that the star might be the source of an interstellar radio signal. In the case of extragalactic distances, however, this potential advantage is less useful because of the increased density of stars per unit area of sky and source confusion, although it could be useful for foreground stars.

### 2.4. HI Spectral Window in Target Velocity Frame

Our observations in the 'HI' spectral window assume that broadcasters aim to catch the attention of astronomers in the Milky Way who are studying neutral hydrogen in the broadcaster's galaxy at high spectral resolution, or aim to catch the attention of SETI observers searching the broadcaster's galaxy for radio signals in the hydrogen band with appropriate Doppler adjustments for that galaxy. In either case, observers in the Milky Way who are constrained by limited spectral windows, as we were, must compensate for the Doppler shift due to systemic velocity of the targets, approximately -300 km s$^{-1}$ in the case of M31 and -200 km s$^{-1}$ for M33, and cover the range of velocity of various parts of the targets due to their rotation—approximately 0 to -600 km s$^{-1}$ for M31 (Brinks & Shane 1984) and -50 to -350 km s$^{-1}$ for M33 (Putman et al. 2009). We covered the velocity range in each field by using 1 MHz windows spanning 200 km s$^{-1}$ and changing center frequency by typically 0.5 MHz or 100 km s$^{-1}$ for each field along the axis of the galaxy, relying on existing velocity maps. With 8192 channels available, a spectral resolution of 122 Hz was possible.

Hydrogen emission in the target galaxy is a potential problem because it increases noise, but at high spectral and spatial resolution it is attenuated and was not a problem in analysis.

### 2.5. LSR Spectral Window, HI in Our Local Standard of Rest

Observations in the 'LSR' spectral window assume that broadcasters aim to catch the attention of observers searching for radio signals in the hydrogen band within our part of the Milky Way when they happen to observe in the direction of the broadcaster's galaxy, or aim to catch the attention of observers searching the target galaxy for radio signals and presuming signals are Doppler-adjusted to our local standard of rest. In either case, Doppler adjustment is needed for our LSR, and the spectral window should cover its uncertainty and the Doppler shifts due to velocity of the Sun and Earth with respect to the LSR. Adjusting for the LSR was done automatically in observing. The uncertainty in the components of the LSR are 2 km s$^{-1}$ or smaller (Schönrich, Binney & Dehnen 2010), and the Sun and Earth velocities together don't exceed about 20 km s$^{-1}$; at 21 cm the Doppler shift is about 4.7 kHz per km s$^{-1}$, so a 100 kHz window is sufficient. We used a spectral window of 125 kHz with 8192 channels resulting in spectral resolution of 15.3 Hz. This LSR spectral window does not usually overlap with the HI window, and is typically free of target galaxy hydrogen emission, although not free of local emission.

Searching for signals that are Doppler-adjusted to our neighborhood of the Milky Way tacitly assumes broadcasts illuminating small areas with high-gain antenna systems Doppler-adjusted to the target areas, but does not assume Doppler-adjustment to our Sun's frame of reference although it includes that possibility.

### 2.6. Wide Spectral Windows at 1450 and 1650 MHz

Brief observations were also made using wide 128 MHz spectral windows centered at 1450 and 1650 MHz with 15.6 kHz resolution, to avoid the assumption of signals based on the 21-cm wavelength, and covering 70% of the so-called waterhole between 1420 and 1665 MHz (Oliver 1977). That analysis is complicated by strong HI emission in small parts of the lower band and by many strong presumably man-made radio signals in the upper band which are commonly seen in VLA surveys (for example Bihr et al. 2016) and has not been completed.

### 2.7. Sensitivity

The theoretical sensitivity of each experimental setup is presented in Table 4. Actual sensitivity varied due to flagged antennas and baselines, processing details, and by channel; it's shown later in Table 5 for features with the highest SNR in each field.

| Spectral window | Int. time (min.) | Spectral window (MHz) | $S_{min}$ all chan. (mJy beam$^{-1}$) | Chan. width (Hz) | $S_{min}$ Chan. (mJy beam$^{-1}$) |
|---|---|---|---|---|---|
| HI | 20 | 1 | 0.37 | 122 | 33.8 |
| LSR | 5 | 0.125 | 2.1 | 15.3 | 190.7 |
| VGR1 | 5 | 16 | 0.1 | 1953 | 10 |

Table 4
Theoretical Sensitivity $S_{min}$

Note: $S_{min}$ calculated using the NRAO exposure calculator ECT for 27 antennas, dual polarization, natural weighting, winter.

## 3. ANALYSIS

### 3.1 Data Reduction and Feature Search

Data were reduced by two different analysts using two largely different software systems—CASA (NRAO 2016) and AIPS (Greisen 2003)—to create clean images, then the AIPS source-finding task SAD ('Search And Destroy') was used to search the resulting >10$^5$ single-channel images for possible radio signals. SAD and other source-finding algorithms have been compared using real data in Mooley et al. 2013; see also Hopkins 2015. Features found in both analyses were viewed as more reliable—less likely to be consequences of choices made in flagging, calibration, and imaging. We use the term 'features' because we were not searching for conventional radio sources and because most were due to noise.

The standard calibrator 3C48 was used as absolute flux, bandpass, and complex gain calibrator for all galaxy observations, located only a few degrees away from M33 and 15 degrees away from M31.

In CASA work, grid and cell sizes were chosen to generate images covering 0.93$^o$ - 0.99$^o$, adjusting for the resolution of various array configurations. In AIPS work, 1$^o$ images were generated on a 2048-pixel grid using a cell size of 1.75" for all fields, without adjusting for the varying resolution of the array. Using a beam model due to Perley given in AIPS tasks such as PBCOR, sensitivity with respect to the center of the field is reduced by a factor of 0.05 at the edge of a 0.93$^o$ beam and 0.02 at the edge of a 1$^o$ field, so the extra spatial coverage results in greatly reduced sensitivity near the edges. The advantage of wider fields is that they cover stars on the periphery of the galaxies and offer a greater opportunity for detecting features in multiple observations.

#### 3.1.1. CASA Data Reduction and Imaging

The raw Jansky VLA data (SDMs) were imported into CASA to produce measurement sets. A custom CASA script was utilized for RFI flagging and using the 3C48 scans for the complex gain, bandpass, and absolute flux density scale calibration of the measurement sets. Flagging of the calibrated target fields (typically 15 baselines or less out of 351 possible) was carried out after inspecting amplitude versus channel plots in CASA task plotms. Channel numbers between 201 to 8000 were then imaged all together interactively with the CASA task clean to make a single "CH0" (deep) image (channels outside this range did not have good bandpass calibration). Each channel within this range was then split off to prepare an independent measurement set for each channel. Each single-channel measurement set was then imaged with the CASA task clean using parameters defined by the CH0 image. Typically, 0.95$^o$ (out to the 5% point of the primary beam) was imaged with ~4 pixels per synthesized beam, and clean boxes defined for the CH0 image were used. Single-channel images were exported to FITS format. The AIPS task RMSD was used to generate the local RMS noise maps for these images, and then fed into feature-finding task SAD to generate 5σ catalogs.

#### 3.1.2. AIPS Data Reduction and Imaging

UV data were initially reviewed and flagged manually, typically excluding a few short baselines with flux greater than twice the typical level and occasionally excluding antennas affected by RFI or other problems. Automated flagging was not used, to avoid inadvertently flagging signals of interest. The task CORER was used to identify baselines with more than a few apparent problems and typically up to about 25 baselines were flagged.

After flagging and calibration, CH0 images were created using the task IMAGR averaging the central 8001 channels (excluding bandpass edges) to get high sensitivity to continuum sources—a factor of 8001$^{1/2}$ better than single channels. Continuum sources were identified in these images by inspection and by using the feature-finding task with a SNR threshold ranging from 7 to 9, accepting features with lower values if they also appeared in NVSS images (Condon et al. 1998). Features were classified as continuum sources if a 1 MHz spectrum displayed no narrow spectral features, and in later analysis of single channel data the area near those coordinates was ignored (typically 60" boxes) in both AIPS and CASA feature finding.

Single-channel images were then created, having much higher sensitivity to narrow bandwidth signals (by a factor of 8001 for coherent signals in a single channel) and much lower sensitivity to continuum sources. The task SAD was then used to identify features in each channel, searching at successively declining flux levels (3, 2, 1.5, 1.25, and 1 times the RMS, times a 4σ threshold, generating catalogs of apparent source fluxes and positions (following Gray & Marvel, 2001). Two hundred channels were usually ignored on band edges after feature finding, increasing to 300 or 400 if the feature with the highest SNR or many of the top 25 features fell within 100 channels of the adjusted band edge, which sometimes occurred due to bandpass calibration problems.

3.2. Statistical Thresholds

After imaging, the 8192-channel galaxy data sets consisted of between $6 \times 10^7$ and $8 \times 10^8$ beams for the various fields, and all fields together totaled $1.69 \times 10^9$ beams (in the CASA analysis). We used a statistical threshold to screen features and identify those with the highest SNR for possible further analysis. To calculate the threshold, we solved the equation:

$$N \, \text{erfc}(SNR/\sqrt{2}) = 1$$

where N is the number of independent beams (image size divided by beamsize, times the number of channels) and SNR is the signal-to-noise ratio, resulting in thresholds ranging from 5.7σ to 5.9σ for various fields. To obtain a threshold above which noise peaks would not be expected, we used SNR+1 (following Frail 2012) and rounded it up to 7.0σ. For all fields combined, the calculated threshold was 6.2σ, and adjusted by SNR+1 was 7.2σ.

3.3. Feature Analysis

Analysis focused on features with the largest SNR in each field, typically in the range 6.8-7σ. First, the spectrum at the feature position was inspected, and if no spectral feature exceeding the local noise peaks was seen, the feature was rejected as spurious—usually confirmed by finding implausible peak/flux or major/minor beam size values in results reported by the feature finding task.

If a spectral feature was apparent, a more carefully cleaned image was created for that channel (using a 'clean box' at the feature's position), and the SNR was calculated using the maximum flux and histogram RMS reported by IMEAN for that channel (rather than global RMS, in the case of AIPS work). The resulting SNR was almost always smaller than the values reported by the feature finding task (which used fitted peak and global RMS in the case of AIPS work) which filtered out some features.

Investigation of top features always included reviewing higher resolution spectra (101 channels to determine spectral feature width) and inspecting NVSS and POSS II images for radio and optical counterparts, and in selected cases included checking match to beam shape, checking right and left polarization, and imaging shorter time ranges.

3.4 VGR1 Detection Example

The Voyager 1 spacecraft was observed as a test, using a JPL HORIZONS ephemeris (Giorgini 2015), and its approximately 10 W signal near 8.4 GHz (Ludwig & Taylor 2002) was easily detected at a range of about 130 AU which is beyond the edge of the solar system. The 'VGR1' detection demonstrates the plausibility of detecting a weak relatively narrow-band radio signal at long range with the VLA, provides an example of tools used, and illustrates some special considerations in SETI (examples are from the AIPS analysis).

Figure 2 shows a needle plot of SNR by channel for all features over 4σ reported by the feature-finding task, with some features approximately 200σ.

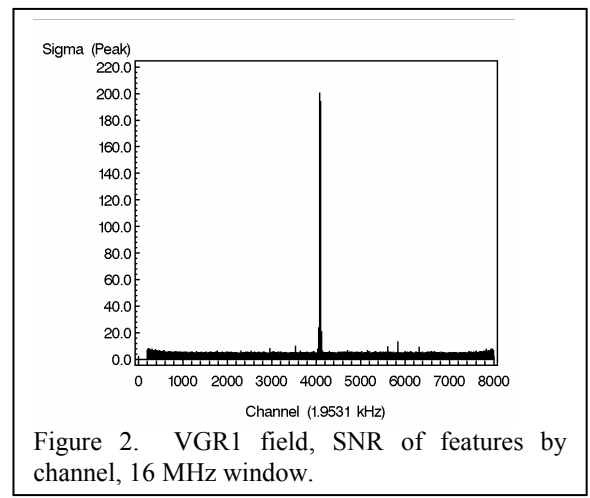

Figure 2. VGR1 field, SNR of features by channel, 16 MHz window.

Figure 3 shows the spectral details in a 101-channel spectrum at the spacecraft position, resolving three strong signals as well as other detail.

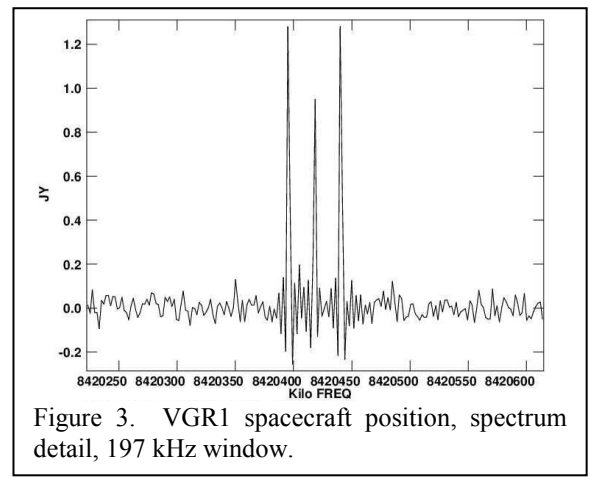

Figure 3. VGR1 spacecraft position, spectrum detail, 197 kHz window.

Figure 4 shows all features by position, with circles scaled to the square of SNR to emphasize large values. The stronger signals all map a single point source at the predicted position for the spacecraft, illustrating spatial correlation as one strength of using synthesis imaging for SETI. The many dots scattered across the field are tiny circles mostly due to noise; some small circles at the VGR1 position are thought to be ringing. A 0.7 Jy continuum source which was prominent in an averaged image does not appear as a feature in single channels, although it was not masked in this test.

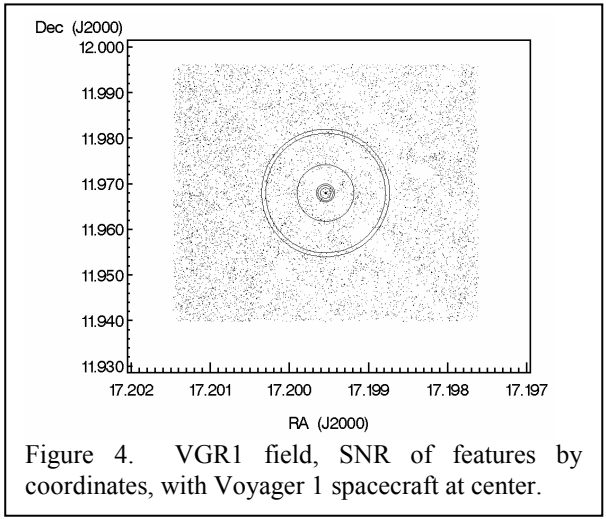

Figure 4. VGR1 field, SNR of features by coordinates, with Voyager 1 spacecraft at center.

Figure 5 shows a map detail of flux (Stokes I) for the single channel 4101 (8.420441346 GHz), one of three channels with strong signal, and the 1.2 Jy flux is prominent with SNR=179. Mapping to a point source which resembles the synthesized beam is evidence of a real signal, which is known to be the case in this test.

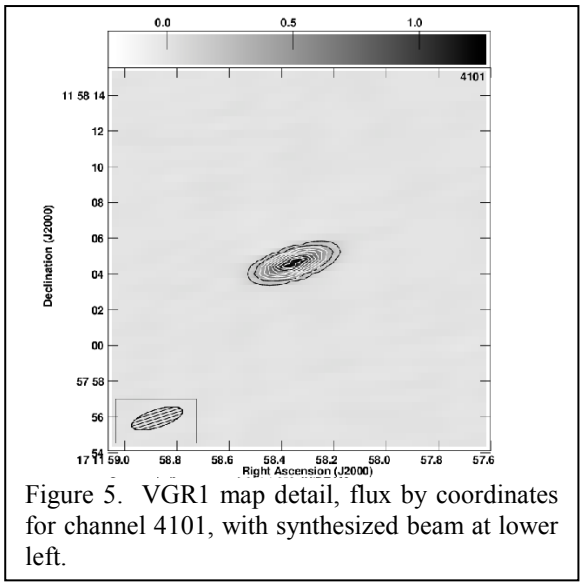

Figure 5. VGR1 map detail, flux by coordinates for channel 4101, with synthesized beam at lower left.

The VGR1 detection illustrates several important aspects of searching for interstellar radio signals.

Radio signals can be highly polarized, and the VGR1 signals were detected only in left polarization. Taking channel 4101 as an example, the task IMEAN reported 1.21 Jy Stokes I total intensity, 2.44 Jy in left polarization, and only 39.6 mJy in right—twice the flux when the telescope polarization matched the signal, compared with the total intensity. Only total intensity was used in most galaxy analysis due to lengthy processing times (16 hours to image 8192 channels in some fields), which could incur a 50% loss for a signal with a single circular polarization during the integration time. Some SETI observations observe and analyze left and right circular polarization separately, often in ~1 Hz channels with ~1 sec. cadence, anticipating the possibility of single-polarization or polarization-switching modulation (Dixon 1973). With our 5 to 20 min. integration times we might incur the loss if a signal used only one circular polarization, but not if polarization varied on a faster time scale.

The importance of narrow channels for detecting narrowband radio signals is illustrated by comparing the channel 4101 flux at the VGR1 position with a wider band consisting of 4005 channels (the approximate upper half of the spectral window, selected to exclude the two other signals), and considering only left polarization which contains signal. The single-channel flux at the VGR1 position was 2.441 Jy, compared with the wideband flux 0.000676 Jy, a factor of 2.441/0.000676=3611 improvement for the narrow channel and 90% of a 4005 factor expected improvement. Failing to achieve 100% of the expected improvement is not surprising because the flux in the wideband average at the VGR1 position was only 4.5 times the 0.150 mJy RMS, which is barely detectable even though a strong narrowband signal was present. This illustrates the advantage of using narrow channels to search for presumably narrowband signals, and it also illustrates the difficulty of detecting narrowband signals with the wide channels often used in radio astronomy.

The fact that the narrowband signal was (barely) detectable in a wide-band average also illustrates the danger of presuming that features detected by averaging many channels are continuum sources, without inspecting their spectra. In this case, the single-channel signal was detectable in the wider band; if it had been classified as a continuum source and its position masked, then a strong narrowband signal could have been missed.

Finally, it is interesting to note that the detection of VGR1 is not just a consequence of averaging for five minutes; it was also detected in single 5-second integrations. For example, the channel 4101 signal in left polarization yielded SNR=212 for the total 265 sec. observation, and SNR=38 in a single 5-sec. integration (not precisely following the expected sqrt(time) decline because the signal strength varies). In principal, analysis could be conducted at the level of 5-sec. integrations for higher sensitivity to transient signals, but we did not do so because of the much greater processing load.

## 4. RESULTS AND DISCUSSION

### 4.1 M31 and M33

The galaxy data were processed and analyzed using the procedures described earlier and illustrated with VGR1; results for the eight fields and two spectral windows on each are tabulated in Table 5.

No feature was found exceeded the 7σ threshold for a single field or 7.2σ for all fields combined, after excluding clearly spurious features. One 7σ feature was reported from the CASA analysis, but the AIPS analysis found only 6.5σ for the position and channel after more careful imaging, and further analysis found no evidence that the feature was interesting. With no features robustly exceeding our statistical threshold, the case is strong to conclude that no signals of interest were present.

In addition to statistical screening, we also investigated top features for evidence of terrestrial Doppler drift, optical counterparts, or other signs of potential interest, discussed later. One example from the galaxy analysis is presented below; results shown are from the CASA analysis unless otherwise noted.

4.2. M31 HI CTR Example

Figure 6 shows features found in a CH0 image for the M31-HI-CTR (center) field, with several sources clearly detected. The strongest was over 177σ (flux density=191 mJy, rms=1.08 mJy), and a total of six features exceeded 6σ. Fluxes were not corrected for primary beam pattern. All other features were below 5.2σ and presumed noise or possibly some weak sources. CH0 features were confirmed as continuum sources by inspecting spectra and NVSS images.

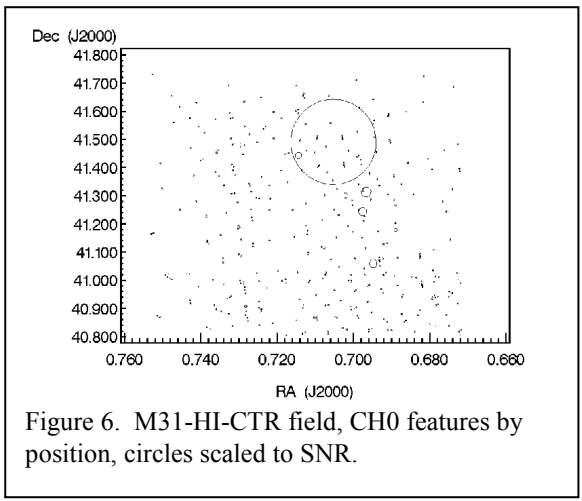

Figure 6. M31-HI-CTR field, CH0 features by position, circles scaled to SNR.

Figure 7 shows the peak SNR by channel for all features in all channels for the field, with continuum sources masked. In this atypical case 1,000 channels were omitted on the high-frequency end of the spectrum due to interference. Two features were above the statistical threshold and the spectra for those of those features were inspected for potentially interesting features. Both features were found to be spurious—not present in spectra or maps, and with implausible beam parameters—and no other features exceeded the threshold.

4.3. Analysis of Spectral 'Features'

The several features with the highest SNR in each field were investigated in more detail, even though they did not exceed the threshold and were presumably due to noise. Features with the highest SNR that appeared in both CASA and AIPS results were given extra attention. The reason for investigating features that are not far above the expected noise peaks is that evidence of real signals might be present, such as the terrestrial Doppler drift signature, optical counterparts, and other signs discussed below.

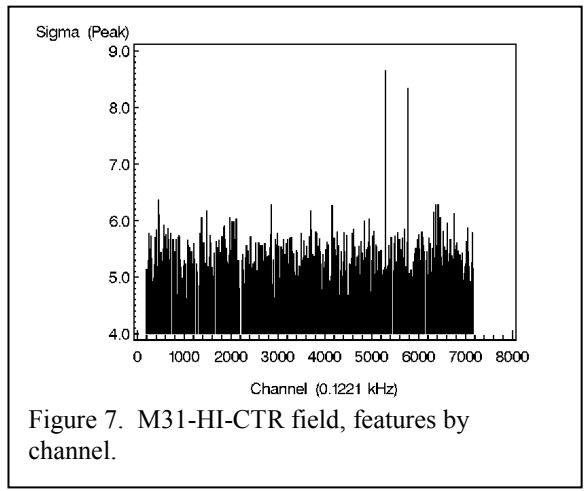

Figure 7. M31-HI-CTR field, features by channel.

*Terrestrial Doppler drift.* Spectral resolution was high enough that a signal drifting with the terrestrial diurnal Doppler rate of up to approximately -7 Hz min$^{-1}$ might drift through several channels during an observation, which could be a 'smoking gun' for an interstellar signal. The drift rate was computed using well-known code (Ball, 1969) and no sign of drift was seen in any of the top features; all were single-channel which is consistent with noise peaks. That drift rate might be difficult to detect, however, because dwell times were chosen in part so that drifting signals would not cross many channels during an observation. In the 5-min. LSR observations with 15-Hz channels, a signal drifting at -7 Hz min$^{-1}$ would drift 35 Hz and might cross two or three channels in the absence of any other effects, and in the 20-min. HI observations with 122 Hz channels, a signal drifting at that rate would drift 140 Hz and might cross one or two channels.

*Optical or radio counterparts.* Spatial resolution was sufficient to potentially identify radio or optical counterparts for features. Finding a radio counterpart for a feature might suggest a continuum source that was not masked, or conceivably a radio source with unexpected spectral lines or flux variation. Finding an optical counterpart for a narrow band feature could indicate an artificial radio signal, which could be very interesting. But, no convincing counterparts for top features were found in either NVSS or POSS, or in searches of astronomical compilations such as SIMBAD. Looking for optical counterparts may be a poor test for stars in the target galaxies because many are unresolved at the long range, but it is a useful test for some foreground stars.

| Table 5 Summary of Observations and Results | | | | | | | | |
|---|---|---|---|---|---|---|---|---|
| Field name | Date | Array config. | Frequency center (GHz) | Velocity center (km s$^{-1}$) | Top Feature | | | |
| | | | | | SNR | Flux density (mJy beam$^{-1}$) | Channel RMS (mJy) | Channel (number) |
| M31-HI-N3 | 2015 Jan 01 | CnB-B | 1.421018 | -150 | 6.3 | 193.2 | 30.73 | 6538 |
| M31-HI-N1 | 2015 Jan 06 | C-CnB | 1.421257 | -200 | **7.0** (6.5) | 232.0 | 33.14 | 4885 |
| M31-HI-CTR | 2015 Jan 27 | CnB-B | 1.421729 | -300 | 6.4 | 188.49 | 29.57 | 0448 |
| M31-HI-S1 | 2015 Jan 06 | C-CnB | 1.422202 | -400 | 6.2 | 206.0 | 33.27 | 1970 |
| M31-HI-S3 | 2015 Jan 06 | C-CnB | 1.422677 | -500 | 6.7 | 231.0 | 34.53 | 1582 |
| M31-LSR-N3 | 2014 Dec 12 | C | 1.420313 | LSR | 6.6 | 1427.0 | 214.93 | 3167 |
| M31-LSR-N1 | 2014 Dec 30 | C | 1.420313 | LSR | 6.5 | 1389.0 | 213.02 | 3568 |
| M31-LSR-CTR | 2014 Dec 30 | C | 1.420313 | LSR | 6.7 | 1295.0 | 194.23 | 6037 |
| M31-LSR-S1 | 2014 Dec 30 | C | 1.420313 | LSR | 6.7 | 1244.0 | 187.15 | 7033 |
| M31-LSR-S3 | 2014 Dec 30 | C | 1.420313 | LSR | 6.5 | 1427.0 | 220.69 | 7812 |
| M33-HI-N1 | 2015 Jan 17 | CnB-B | 1.421451 | -250 | 6.4 | 222.0 | 34.73 | 4191 |
| M33-HI-CTR | 2015 Jan 6 | C-CnB | 1.421221 | -200 | 6.6 | 214.0 | 32.31 | 2856 |
| M33-HI-S1 | 2015 Jan 6 | CnB-B | 1.420984 | -150 | 6.7 | 209.0 | 31.21 | 7021 |
| M33-LSR-N1 | 2014 Dec 22 | C | 1.420288 | LSR | 6.3 | 1166.0 | 185.94 | 6230 |
| M33-LSR-CTR | 2014 Dec 22 | C | 1.420288 | LSR | 6.1 | 1241.0 | 202.80 | 3865 |
| M33-LSR-S1 | 2014 Dec 22 | C | 1.420288 | LSR | 6.7 | 1463.0 | 216.95 | 5289 |
| VGR1 | 2015 Jan 17 | CnB | 8.420432 | ~0 | 200.7 | 1361.0 | 6.78 | 4078 |

Note: Galaxy results are from CASA analysis using RMS for each channel; VGR1 results are from AIPS analysis using global RMS. Bold indicates feature SNR meeting or exceeding threshold; in the case of M31-HI-N1 the smaller AIPS SNR is shown in parenthesis.

*Spatial correlation*. Multiple spectral features at a common position could indicate a signal wider than a single channel, or Doppler drift of a narrow signal over adjacent channels, or a polychromatic comb of signals (Cohen & Charlton 1995), or other effects. We searched for features with approximately common coordinates but different channels, using SQL to calculate the distance between all feature positions, for both the top 25 features in each field based on SNR, and for larger samples in each field, and across all fields for each galaxy. Few features were within a few arcsec of others, and none that were within a few channels had any other unusual characteristics.

*Multiple detections*. Some fields overlapped, so most parts of the target galaxies were observed at least twice, so multiple detections were possible and could provide evidence of a real source below the statistical threshold. Overlap in frequency was only partial for the HI spectral window, but it was complete for the LSR window. The HI 1-MHz window covered 200 km s$^{-1}$ (less was useful due to excluded bandpass edges), and tuning was typically changed by 100 km s$^{-1}$ or 0.5 MHz between adjacent fields, so somewhat less than half of the same spectral window was observed in multiple observations. Pooling all fields for a galaxy, none of the top features had the same position and frequency, or adjacent channels suggesting Doppler drift.

*Present in both CASA and AIPS analysis*. Ten 'features' out of the top approximately 50 (based on SNR) appeared in both the CASA and AIPS analysis at the same position and channel, in the same field. But, the underlying data were the same, not two independent observations, so finding the same features in both was not a multiple detection. These features were investigated more carefully using the methods above, and no evidence was found that they were anything more than noise peaks or possibly RFI or imaging artifacts.

*Polarization*. Evidence of something unusual might be high or total polarization, and for selected features both RCP and LCP were imaged separately and inspected. Separate polarizations were not inspected for all fields or all features.

4.4. M33-LSR-S1 Channel 5289 Example

One 'feature' is described below—emphatically not as a candidate interstellar radio signal—but as an example of a feature near the thresholds and as an example of typical investigation of such features, which found no evidence to suggest anything other than noise.

The top feature for the M33-LSR-S1 field was a 6.75σ feature in the CASA analysis, in channel 5289 (1420.306438 MHz) at RA 1$^h$ 32$^m$ 53.64$^s$, Decl. 30$^o$ 14' 34.40" (flux density=1463 mJy, rms=217 mJy) which somewhat exceeds the 6.7σ threshold computed for the field.

The feature appeared fairly prominent in a 125-kHz spectrum for that position—flux more than twice that of most other peaks, and well above five times the channel RMS—shown in Figure 8.

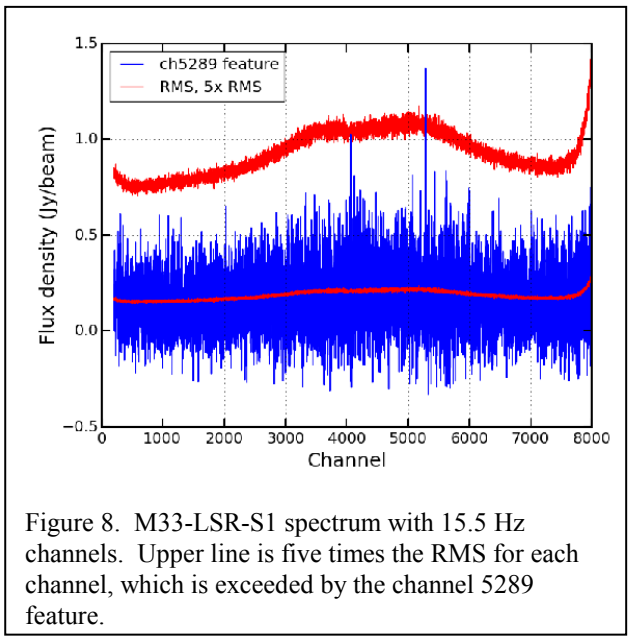

Figure 8. M33-LSR-S1 spectrum with 15.5 Hz channels. Upper line is five times the RMS for each channel, which is exceeded by the channel 5289 feature.

The AIPS analysis for that field found a $8.01\sigma$ feature in the same channel and essentially same position RA $1^h\ 32^m\ 53.63^s$, Decl. $30^o\ 14'\ 34.91"$, also as the top feature in the field. With more careful cleaning, however, the SNR declined to $6.57\sigma$ (flux density=1630 mJy, rms=248 mJy) making the feature drop below our threshold and suggesting a noise origin.

The feature appeared beamlike in clean maps and did not display any of the symptoms of spurious features. Most of the tests described earlier were applied, none suggesting evidence that it was anything more than a statistical peak due to noise or instrumental effects. No evidence of Doppler drift was found; the feature appeared in only one 15.3-Hz channel. No optical counterpart was found, and one might have been resolved if the source was a foreground star. This M33-S1 field position and frequency was also covered by the M33-CTR field, but no feature was seen there. The feature appeared in both right and left circular polarization at 1.6 and 1.8 Jy respectively, and several other features of approximately equal flux density at other positions in the same channel were seen in a RCP map which made this feature not unique and therefore less interesting.

Failing to find any interesting properties leads us to conclude that this was unlikely to be a real narrowband radio source.

4.5. Constraints on Signals

No features were detected exceeding our $7\sigma$ threshold, which in units of flux density was approximately 0.24 Jy for 122-Hz channels, and 1.33 Jy for 15-Hz channels, based on the $1\sigma$ typical RMS sensitivity for each spectral window shown in Table 5.

This constrains the power of hypothetical constant transmissions from M31 or M33 to $\sim 10^{17}$ W if illuminating the entire Milky Way, or as little as $\sim 10^{13}$ W from a 1000-m scale antenna system if it was present when we looked.

4.6. Future Work

The attraction of searching $\sim 10^{12}$ stars in a relatively brief time seems sufficient to search a much wider spectral window than the relatively narrow 125 kHz and 1 MHz windows we observed, and for more than our 5 or 20 minutes. With the approximate direction of hypothetical broadcasters assumed 'known', major unknowns remaining are frequency, duty cycle, and flux.

Signals might be broadcast at frequencies other than the hydrogen-based spectral windows we investigated, or be offset to fall in other velocity frames such as the CMB (Kogut et al. 1993) or in our galactic center of rest (Dixon 1973). Increasing capabilities of spectrometers allow increasingly wide spectral windows to be searched. For example, the Allen Telescope Array can observe 100 million ~1-Hz channels in several 3.9' beams in a phased array mode (DeBoer 2004; Tarter 2011; Welch 2009), SERENDIP V.v covers 300 MHz at 1.49 Hz (Siemion et al. 2011), and *Breakthrough Listen* proposes ~1 Hz spectral resolution over ~10 GHz (Merali 2015).

Transient signals seem especially worth considering in the context of long-range searches such as galaxies, because average power requirements could be reduced by many orders of magnitude by reducing transmission duty cycles. To find intermittent signals, we would need to dwell for unknown and possibly extended periods of time, but such signals might repeat—hopefully at some 'reasonable' rate such as planetary days, and perhaps periodically. For example, we could monitor $\sim 10^{12}$ stars in M31 for approximately 24 hours by observing the five fields used here for 24 hours each with the VLA, yielding an eight-fold increase in sensitivity and taking a total of about one week including overhead. The resulting data could be averaged over many hours searching for weak continuous signals, and over short time ranges searching for transient signals.

Figure 9 shows this survey in the context of selected prior SETI observations, drawn from a comprehensive list (Tarter 1995 and updates) which is fairly complete until 2012, and includes most of the benchmark searches excluding pulsar searches.

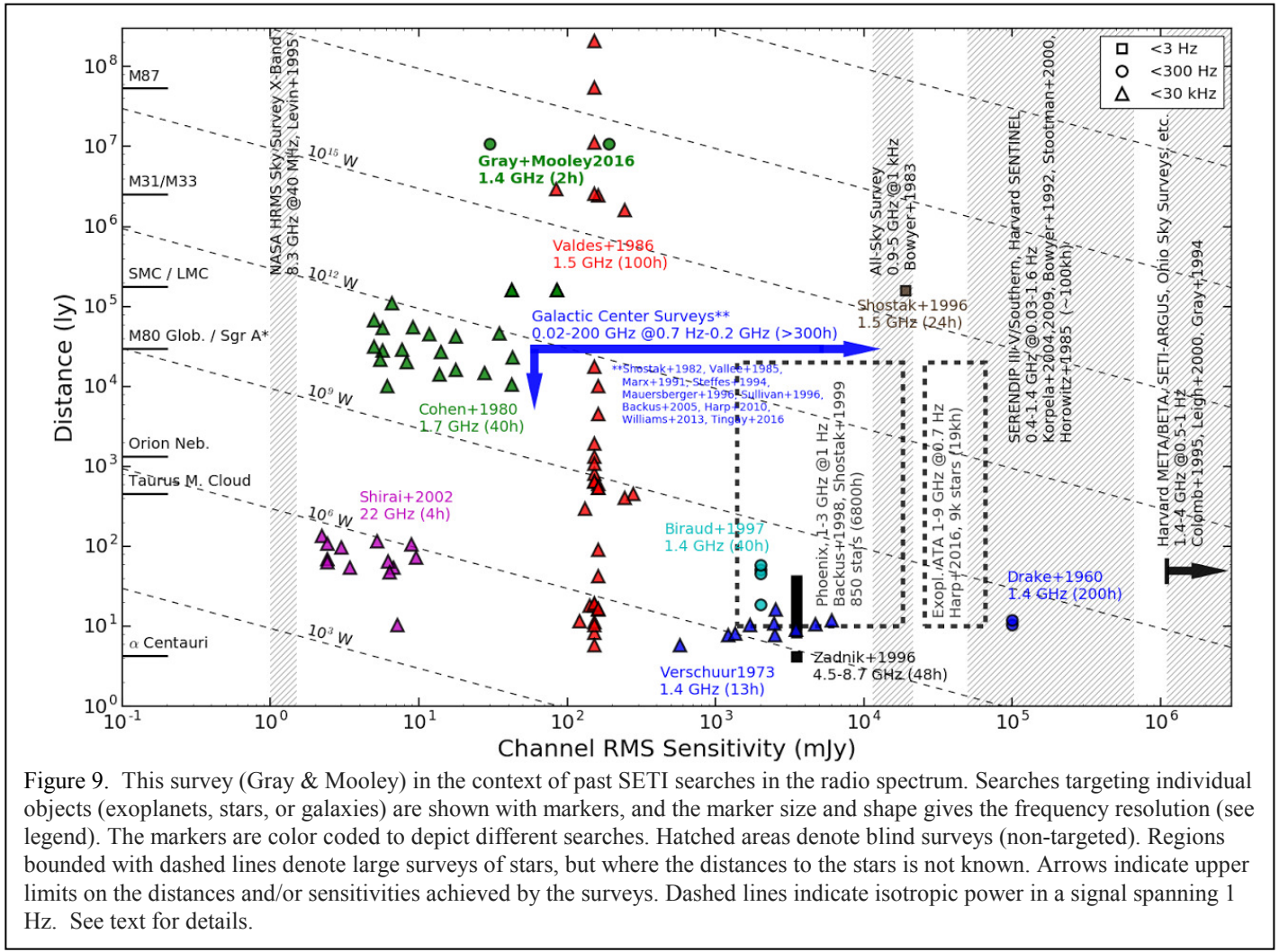

Figure 9. This survey (Gray & Mooley) in the context of past SETI searches in the radio spectrum. Searches targeting individual objects (exoplanets, stars, or galaxies) are shown with markers, and the marker size and shape gives the frequency resolution (see legend). The markers are color coded to depict different searches. Hatched areas denote blind surveys (non-targeted). Regions bounded with dashed lines denote large surveys of stars, but where the distances to the stars is not known. Arrows indicate upper limits on the distances and/or sensitivities achieved by the surveys. Dashed lines indicate isotropic power in a signal spanning 1 Hz. See text for details.

## CONCLUSIONS

No obvious radio signals were detected in JVLA 21-cm band observations of M31 and M33, in either 1 MHz spectral windows centered on the target galaxy rest frame dwelling for 20 min. with 122-Hz channels, or in 0.125 MHz spectral windows centered on our LSR dwelling for 5 min. with 15-Hz channels, above a signal-to-noise ratio of seven. This constrains putative constant emissions to approximately 0.24 Jy in the case of the 122-Hz channels, and 1.33 Jy in the case of the 15-Hz channels. This does not rule out the possibility of signals that are much briefer than our dwell time, or much longer repetition rates, or signals outside of the spectral windows observed, or signals below our detection limit.

This search of two nearby galaxies is significant because it sampled concentrations of many billions of stars, a search strategy previously used by Sagan and Drake in the 1970s for M33 at 1 kHz resolution but not for the ~100 times larger M31, and could have detected a bright continuous 21-cm beacon if one was illuminating the Milky Way.

The absence of detectable signals in a relatively brief search of narrow spectral windows is not sufficient to conclude that no signals exist anywhere in the spectrum amid the many stars of these two prominent Local Group galaxies, but it does demonstrate that there is no "low hanging fruit" such as a bright continuous beacon signal at 21 cm, which had been a possibility.

Observing $\sim 10^{12}$ stars at relatively high spectral resolution (~100 Hz) with a sensitive radio telescope makes this arguably the largest SETI experiment ever reported in terms of potential signal sources, although the distance to the two galaxies would require transmissions of great power. Future searches of M31 and M33 over wider spectral windows with higher spectral resolution and for longer dwell times are feasible, and seem worthwhile.

## ACKNOWLEDGMENTS

RG thanks Patrick Palmer for guidance in many aspects of this work, Elias Brinks for helpful comments during planning, and Walter Brisken, Eric Greisen, Steven Lord, Heidi Medlin, Drew Medlin, and Juergen Ott for valuable advice. An anonymous referee provided helpful critique. Observations were made under project code VLA-14B-292. This work was supported in part by the SAS Institute with software used for much of the analysis and graphics. KPM's research is supported by the Oxford Centre for Astrophysical Surveys, which is funded through generous support from the Hintze Family Charitable Foundation

The National Radio Astronomy Observatory is a facility of the National Science Foundation operated under cooperative agreement by Associated Universities, Inc. This research made use of "Aladin sky atlas" developed at CDS, Strasbourg Observatory, France; the SIMBAD database, operated at CDS, Strasbourg, France; and the NASA/IPAC Extragalactic Database (NED) which is operated by the Jet Propulsion Laboratory, California Institute of Technology, under contract with the National Aeronautics and Space Administration.